\documentclass[aps,twocolumn,preprintnumbers]{revtex4}

\usepackage{graphicx}  % Include figure files
\usepackage{subfigure}
\usepackage{multirow}
\usepackage{dsfont}

\usepackage{fancyhdr}
\usepackage{longtable}
\usepackage{parskip}
\usepackage[T1]{fontenc}
\usepackage{dcolumn}   % Align table columns on decimal point
\graphicspath{{Images/}}
\usepackage{bm}        % bold math
\usepackage{amsfonts}  % Common math fonts
\usepackage{amsmath}   % Common math functions
\usepackage{amssymb}   % Common math symbols
\usepackage{xcolor}
\usepackage{mathrsfs}
\usepackage{relsize}
\usepackage{hyperref}
\usepackage{mathtools}
\usepackage{natbib}
\usepackage{lipsum}

\begin{document}

\title{Complex Heat Capacity as A Witness of Spatio-Temporal Entanglement}

\author{Mia Stamatova and Vlatko Vedral}

\affiliation {Clarendon Laboratory, University of Oxford, Parks Road, Oxford OX1 3PU, United Kingdom}

\date{August 21, 2025}

\begin{abstract}  
We propose a novel witness of temporal quantum entanglement using the imaginary component of the complex heat capacity - a measurable thermodynamic quantity in temperature-modulated calorimetry. By establishing a direct correspondence between complex heat capacity and the pseudo-density matrix formalism, our approach enables the characterisation of both spatial and temporal quantum correlations without demanding additional state-level manipulation beyond initial tomography. We analytically demonstrate this connection for an open quantum system modelled by a qubit coupled to a thermal bath, and show how both pseudo-density matrix negativity and violations of a temporal CHSH inequality emerge as indicators of non-classical temporal correlations. We further identify bounds on the imaginary heat capacity that guarantee temporal entanglement, providing an experimentally accessible criterion at the macroscopic scale. This framework offers a feasible route for probing temporal quantum effects in condensed-matter systems and opens a viable path toward experimental realisation.
\end{abstract}

\maketitle 

\textit{Introduction}— The treatment of space and time differs vastly in quantum mechanics, both mathematically and conceptually. In non-relativistic quantum mechanics, position is treated as an operator, while time enters as a real-valued, independent background parameter. This asymmetry is further revealed through the lack of a temporal tensor product structure - an important mathematical feature that gives rise to the notion of entanglement in space \cite{nielsen2010quantum}. Nevertheless, even in the absence of an analogous mathematical construction, quantum mechanics predicts non-classical temporal correlations that can be probed via temporal Bell- or Leggett-Garg inequalities \cite{leggett1985quantum, emary2013leggett, brukner2004quantum}.  A violation of these inequalities would imply that the correlations between outcomes of sequential measurements on a single quantum system cannot be explained by any classical hidden-variable model that satisfies macrorealism and non-invasive measurability; and hence, is an indicator of genuine temporal quantum correlations.  

However, experimentally testing for non-classicality —both spatial and temporal—  using this method is difficult when applied to macroscopic bodies, and often impossible due to the inaccessibility of individual subsystem states in a complex system. Alternative witnesses of spatial entanglement that circumvent this problem have been developed in quantum thermodynamics. Namely, previous work \cite{wiesniak2008heat, brukner2004macroscopic, toth2005entanglement, wu2005entanglement} has demonstrated that thermodynamic observables, such as heat capacity and magnetic susceptibility, may be used to test for non-classical spatial correlations. Measuring thermodynamic observables for macroscopic systems is a standard experimental procedure, as opposed to making sophisticated measurements that circumvent various loopholes to violate Bell's inequalities. Thus, this approach enables entanglement to be probed at the macroscopic scale, offering insight into the compelling question of whether quantum correlations at the microscopic level can give rise to observable large-scale effects.

In this work, we extend these ideas to the temporal domain. We show that the imaginary component of the complex heat capacity can be used as a witness of temporal non-classicality, and establish a connection between measurements of the real and imaginary components of the heat capacity to the entries of a pseudo-density matrix (PDM). The PDM formalism \cite{fitzsimons2015quantum, fullwood2025quantum, marletto2019theoretical, marletto2021temporal, zhang2020quantum},  was introduced to allow spatial- and temporal quantum correlations to be placed on an equal footing. They are usually formulated by considering sequential measurements on a single quantum system, where each measurement is made in all three complementary bases $X, Y,$ and $Z$ (represented by the Pauli operators). Applying the Pauli operators to a system requires a precise, coherent control over the system's quantum state, which is a highly non-trivial task for macroscopic bodies due to decoherence effects. This work offers a novel approach to constructing pseudo-density matrices and extends their applicability to the emergent scale. 

Further to this, using the PDM framework, we derive a temporal CHSH inequality \cite{brukner2004quantum, seevinck2002bell} for a single qubit measured at two distinct times, and use it to show that only a subset of values of the imaginary component of the heat capacity yields a violation of a classical bound. This provides a robust criterion for temporal entanglement in the context of quantum thermodynamics, and the experimental feasibility of this approach will be explored in a subsequent follow-up paper.

\textit{A witness of temporal non-classicality}— 
Complex heat capacity is usually studied in the context of temperature-modulated calorimetry \cite{baur1998complex, merzlyakov1999complex, de2019complex, hatta1999some, minakov1999dynamic, garden2007simple, fiore2019entropy}, and arises when a weak temperature modulation is applied to a system, resulting in a sinusoidal time-varying heat flux and naturally leads to a definition of complex heat capacity, $C(\omega) =C'(\omega) + i C''(\omega)$. The physical meaning of the imaginary part of the heat capacity has previously been related to the rate of entropy production \cite{de2019complex, jeong1997progress}, though a link to quantities arising from quantum information theory has yet to be made. Here, we directly relate the imaginary component of the heat capacity to temporal non-classicality.

Consider a macroscopic system in thermal equilibrium. Applying a small time-dependent temperature perturbation $T(t) = T + \delta T(t)$ will result in a response of the Hamiltonian $H(t)$, which we seek to characterise. Assuming the temperature perturbation is small enough that we remain in the linear response regime, the fluctuation-dissipation theorem \cite{callen1951irreversibility,weiss2012quantum} can be used to approximate the perturbation to the Hamiltonian as,

\begin{equation}
    \delta \langle H(t) \rangle = \int_{0}^{t} \chi (t-t') \delta T(t') dt',
\end{equation}
where $\chi$ is the response function, which is obtained via the Kubo formula \cite{kubo1957statistical}, and is given by

\begin{equation}
    \chi = \frac{i}{\hbar} \Theta(t) \langle[H(t), H(0)]\rangle.
\end{equation}
Defining the complex heat capacity as the Fourier transform of the response function, we arrive at the following expression for the imaginary component of the heat capacity
\begin{equation}
 C''(\omega ) = \frac{1}{2 \hbar} \int_{-\infty}^{\infty}dt \langle[H(t), H(0)]\rangle e^{i \omega t}.
\end{equation}
The commutator in the integral implies that if the imaginary component of the complex heat capacity is non-zero, then temporal non-classicality may be present. We will later place a stricter bound on the values of $C''(\omega)$, for which temporal quantum correlations are manifest in an open system consisting of a qubit coupled to a thermal bath, whose imaginary component is given by
\begin{equation}
    C''(\omega) = \left( \frac{(\hbar \omega)^2}{4k_B T^2}\right)\frac{\omega \Gamma}{\Gamma^2 +\omega^2}\text{sech}^2 \left(\frac{\hbar \omega}{2 k_B T}\right).
\end{equation}
For a full derivation, see Appendix.

\textit{Pseudo-density matrices}—The PDM generalizes the density matrix to describe quantum systems evolving across both space and time  \cite{fitzsimons2015quantum, fullwood2025quantum, marletto2019theoretical, marletto2021temporal, zhang2020quantum}. Unlike the conventional formalism, which encodes spatial correlations at a fixed instant, the PDM incorporates temporal sequences of measurements, allowing it to represent the behaviour of a system across multiple points in time. This approach stems from viewing the density matrix not just as a statistical mixture of pure states, but as an object defined by the expectation values of Pauli operator combinations. Formally, it is expressed as 
\begin{equation}
    R = \frac{1}{2^n} \sum_{i_1 = 0}^{3} \dots \sum_{i_n = 0}^{3} \langle \{ \sigma_{i_j}\}_{j=1}^{n} \rangle \bigotimes_{j = 1}^{n} \sigma_{i_j},
\end{equation}
where $\langle \{ \sigma_{i_j}\}_{j=1}^{n} \rangle$ is the expectation value for the product of a set of Pauli measurements \cite{marletto2021temporal}. Note that, although defined in terms of qubits, the framework is extendable to quantum systems of any dimension by embedding the system within a qubit Hilbert space and limiting its dynamics to the relevant subspace. 

Nevertheless, experimental realisation is challenging, as sequential measurements must preserve temporal coherence. Previous work has focused on using photonic systems to simulate temporal PDMs and demonstrate multipartite CHSH violations \cite{marletto2021temporal}; also, NMR experiments have observed temporal correlations via PDM negativity and their decay under decoherence \cite{fitzsimons2015quantum}. These methods, however, are hindered by tomography requirements, coherence loss, and the difficulty of simulating non-invasive sequential measurements without state collapse. As an alternative, we propose constructing PDMs through measurements of the complex heat capacity - a routine solid-state technique that avoids these limitations. 

\textit{Relating heat capacity to PDMs}—Here we relate the matrix elements of the PDM to the complex heat capacity, such that 
\begin{equation}
    R = \frac{1}{4} \sum_{\mu, \nu = 0}^{3} C_{\mu,\nu}(\omega) \sigma_\mu \otimes \sigma_\nu,
\end{equation}
and compute the entries based on the correlation functions of the Hamiltonian components. 

Let $H = \frac{\hbar}{2} \vec{\omega}(t) \cdot \vec{\sigma}$ and assume weak periodic driving is of the form $\vec{\omega}(t) = \vec{\omega}(0) + \delta \vec{\omega} e^{-i\omega t}$. Previously, we used a commutator-based linear definition for heat capacity, 
\begin{equation}
    C_{\mu \nu}(\omega) = \frac{1}{k_B T^2} \int_{0}^{\infty} dt \langle [H_\mu(t), H_\nu(0)]\rangle e^{i \omega t}
\end{equation}
but now demonstrate that it can equivalently be expressed via correlation functions, matching the pseudo-density matrix construction. 
\begin{figure}
    \includegraphics[width=0.49\textwidth]{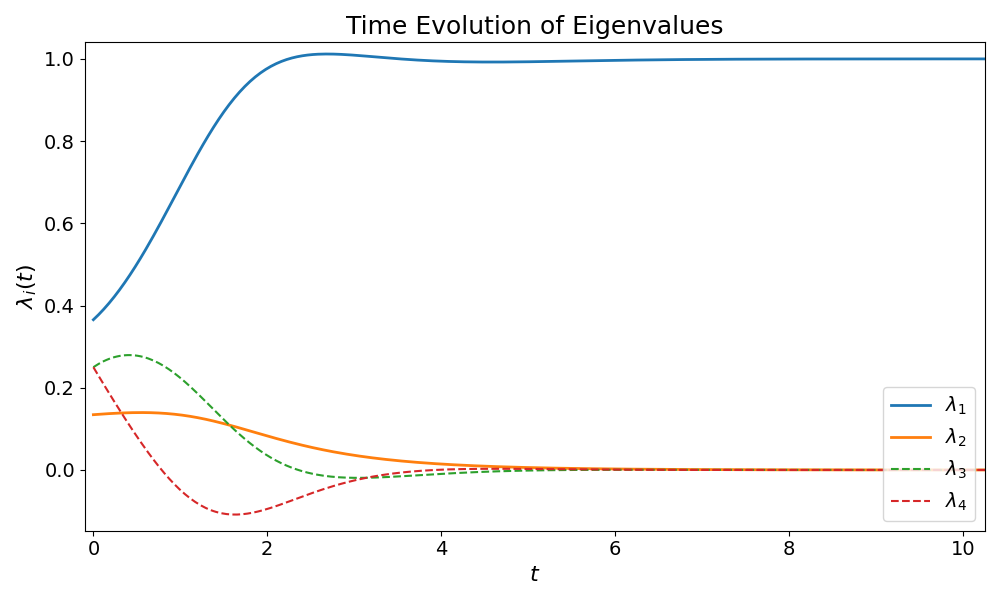}
    \caption{The time evolution of the eigenvalues of a two-time PDM of a single qubit coupled to a thermal bath. Natural units have been adopted, such that: $\hbar = 1$; $k_B = 1$; and we set $\omega_z = 1$; T $=1$; $\gamma_1 = 0.9$; $\gamma_2 = 1.2$.}
    \label{fig:my_label}
\end{figure}
From the periodic driving, the Hamiltonian varies as $\dot{H}(t) = \frac{-i \hbar \omega}{2} \delta \vec{\omega}\cdot \vec{\sigma} e^{-i \omega t} = \Sigma_{\mu = 1}^{3} \frac{-i \hbar \omega}{2} \delta \omega_\mu e^{-i \omega t} \sigma_\mu$. Hence, we can write $H(t) = \Sigma_\mu H_\mu (t) \sigma_\mu$, where $\sigma_\mu$ obeys the usual commutation relation $[\sigma_\mu , \sigma_\nu] = 2i \epsilon_{\mu \nu \lambda} \sigma_\lambda$. Relating this to Eq.(7), it can be shown that $\langle [H(t), H(0)]\rangle = 2i \Sigma_{\mu, \nu, \lambda} \epsilon_{\mu \nu \lambda} \langle H_\mu (t) H_\nu (0) \rangle$, such that 
\begin{equation}
    C_{\mu\nu} (\omega) = \frac{1}{k_B T^2} \int_{0}^{\infty} dt \langle \dot{H}_\mu (t) \dot{H}_\nu (0)\rangle e^{i \omega t},
\end{equation}
whose form allows for the entries of the PDM to be expressed via the complex heat capacity.

To illustrate this idea, let us consider a qubit undergoing dephasing governed by Lindblad dynamics, and explicitly construct the PDM. The Hamiltonian is given by $H = \frac{\hbar \omega_z}{2}\sigma_z$; and let the relaxation rate be denoted by $\gamma_1$, and the overall dephasing rate by $\gamma_2 = \frac{\gamma_1}{2}$ + $\gamma_\varphi$, where $\gamma_\varphi$ is the pure dephasing rate. The system evolves under
\begin{equation}
    \frac{d\rho}{dt} = \frac{-i}{\hbar}[H, \rho] + L_{\text{relax}}[\rho] + L_{\text{dephase}}[\rho],
\end{equation}
where
 $L_{\text{relax}}[\rho]= \frac{\gamma_1}{2} ((n_{\text{th}}+1)(2\sigma_- \rho \sigma_+ - \{\sigma_+ \sigma_- , \rho \})$ + $n_\text{th}(2\sigma_+ \rho \sigma_- - \{\sigma_- \sigma_+ , \rho \}))$ ; and $L_\text{dephase}[\rho] = \frac{\gamma_\varphi}{2}(\sigma_z \rho \sigma_z - 1)$. Note, here $n_{\text{th}}$ has the usual thermodynamic definition,  $n_\text{th} = \frac{1}{e^{\beta \hbar \omega} - 1}$.

 To construct the PDM, we assume that the system has equilibrated at both times $t_0$ and $t_1$, such that $R_{0z} = R_{z0} = \langle \sigma_z(t_0) \rangle = \langle \sigma_z (t_1) \rangle = - \tanh \left( \frac{\hbar \omega_z}{2 k_B T}\right)$. The remaining entries can be calculated via Eq.(5) to obtain the full PDM

\begin{widetext}
\begin{equation*}
R = \frac{1}{4N(t)}
\begin{pmatrix}
    1 & 0 & 0 & -\text{tanh}\left(\frac{\hbar \omega_z}{2 k_B T} \right)\\
    0 & e^{-\gamma_2 t}\cos (\omega_z t) & -ie^{-\gamma_2 t}\sin (\omega_z t) & 0 \\
    0 & ie^{-\gamma_2 t}\sin (\omega_z t) & e^{-\gamma_2 t}\cos (\omega_z t) & 0 \\
    -\text{tanh}\left(\frac{\hbar \omega_z}{2 k_B T}\right) & 0 & 0 & \text{tanh}^2\left( \frac{\hbar \omega_z}{2 k_B T} \right) + \text{sech}^2\left( \frac{\hbar \omega_z}{2 k_B T}\right)e^{-\gamma_1 t}
\end{pmatrix},
\end{equation*}
\end{widetext}
where 
\begin{align*}
    N(t) = \frac{1}{4} &\Bigl(1 + 2 e^{-\gamma_2 t} \cos(\omega_z t) + \tanh^2\left(\frac{\hbar\omega_z}{2k_B T} \right)  \\
    & + \text{sech}^2 \left( \frac{\hbar \omega_z}{2 k_B T}\right) e^{-\gamma_1 t} \Bigr),
\end{align*}
ensuring the PDM is normalised for all times. Physically, the off-diagonal terms arise due to dephasing of the qubit; and the $R_{zz}$ term corresponds to the real heat capacity plus the decaying imaginary component.
 \begin{figure}
    \includegraphics[width=0.49\textwidth]{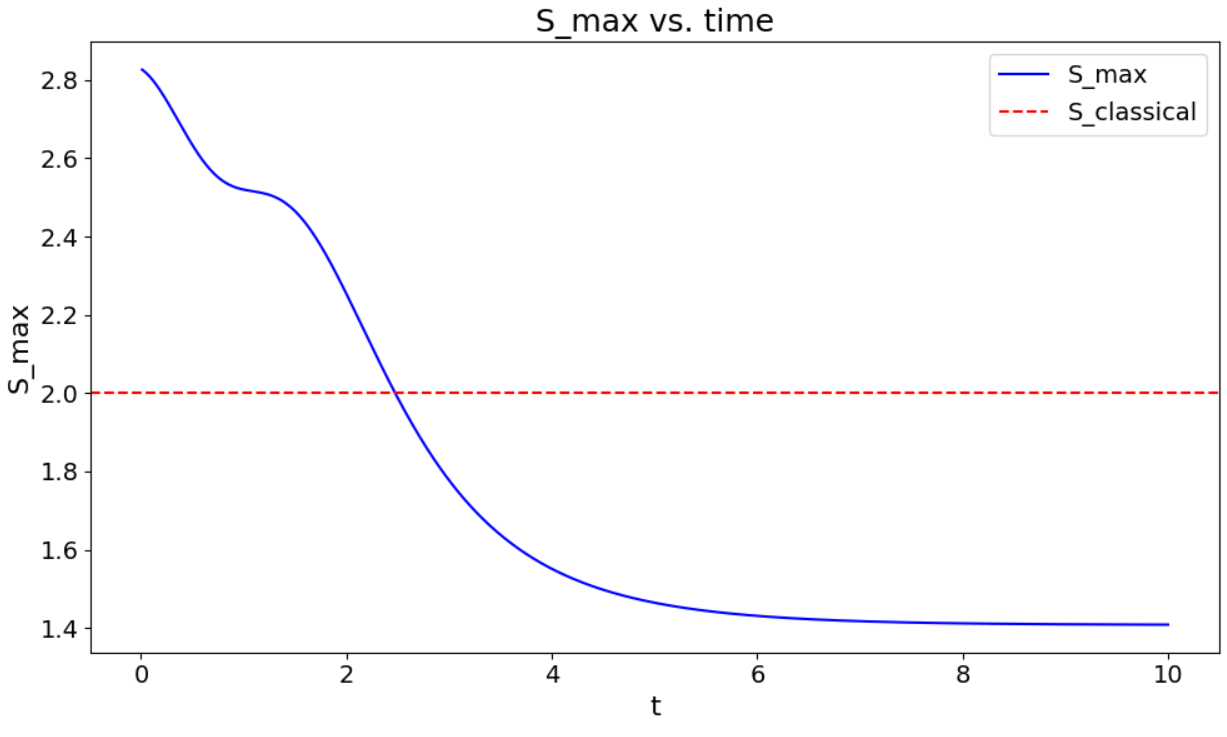}
    \caption{The time evolution of the maximised Bell parameter of a two-time PDM of a single qubit coupled to a thermal bath. Natural units have been adopted, such that: $\hbar = 1$; $k_B = 1$; and we set $\omega_z = 1$; T $=1$; $\gamma_1 = 0.9$; $\gamma_2 = 1.2$.}
    \label{fig:my_label}
\end{figure}
\begin{figure}
    \includegraphics[width=0.49\textwidth]{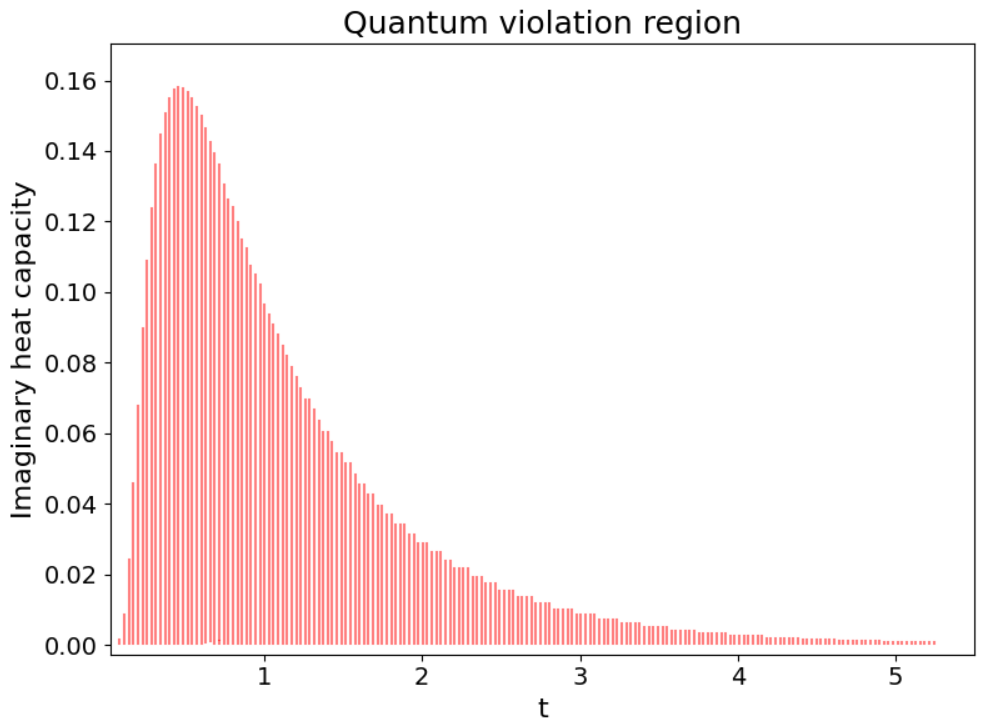}
    \caption{The shaded region shows violation of the temporal CHSH inequality. Temporal entanglement is witnessed only for values of the imaginary component of the complex heat capacity that lie in this region. Natural units have been adopted, such that: $\hbar = 1$; $k_B = 1$; and we set $\omega_z = 1$; T $=1$; $\gamma_1 = 0.9$; $\gamma_2 = 1.2$.}
    \label{fig:my_label}
\end{figure}
The eigenvalues are 
\begin{equation*}
\begin{aligned}
& \lambda_{1,2} = \frac{1}{8 N(t)}\left( 1+ Z \pm \sqrt{(1- Z)^2 + 4 \tanh^2\left( \frac{\hbar \omega_z}{2 k_B T}\right)}\right)  \\
& \lambda_{3,4}= \frac{1}{4 N(t)}\left( \cos(\omega_z t) \pm \sin(\omega_z t)\right) e^{-\gamma_2 t}, 
\end{aligned}
\end{equation*}
where $Z = \tanh^2\left( \frac{\hbar \omega_z}{2k_B T}\right) + \text{sech}^2\left( \frac{\hbar \omega_z}{2 k_B T}\right) e^{-\gamma_1 t}$. Plotting the eigenvalues against time (see Fig. 1), we see that either $\lambda _3$ or $\lambda_4$ is negative - an indicator of temporal non-classicality.  For later times, the eigenvalues all tend to positive values, because the qubit eventually settles into a steady-state that approximately follows the thermal distribution, assuming the temperature modulation is slow.

\textit{A connection to the temporal CHSH inequality}— While the negativity of the eigenvalue does indeed imply the existence of temporal non-classicality, a less contentious criterion (that is, more widely-accepted in the physics community currently) is the violation of a classical bound such as the temporal CHSH inequality. We will now derive this bound in the context of heat capacity.

Consider two sequential measurements on a qubit, where $A_{t_1}^1$ and $A_{t_1}^2$ denote outcomes of dichotomous observables $\vec{a}_1$ and $\vec{a}_2$ at time $t_1$; and $B_{t_2}^1$ and $B_{t_2}^2$ those of $\vec{b}_1$ and $\vec{b}_2$ at $t_2 > t_1$. Predetermined outcomes satisfy $A_{t_1}^1 \left(B_{t_2}^1 + B_{t_2}^2\right) + A_{t_1}^2 \left( B_{t_2}^1 - B_{t_2}^2\right) = \pm 2$, which, upon averaging, yields the temporal CHSH inequality
\begin{equation*}
    S \equiv | E( \vec{a}_1, \vec{b}_1) + E( \vec{a}_1, \vec{b}_2) + E( \vec{a}_2, \vec{b}_1) - E( \vec{a}_2, \vec{b}_2) | \leq 2. 
\end{equation*}
For a more detailed derivation, see \cite{brukner2004quantum}. For arbitrary measurement directions, the correlators are given by $E(\vec{a},\vec{b}) = \vec{a}^T T \vec{b}$, where $T_{ij} = \text{Tr}[R \cdot\sigma_i \otimes \sigma_j]$ defines the $3\times 3$ correlation matrix. Optimisation over unit vectors yields $S_{\max}(t) = 2 \sqrt{\lambda_1 + \lambda_2}$, where $\lambda_1$ and $\lambda_2$ are the largest eigenvalues of $T^{T}T$. A quantum violation occurs when
\begin{equation*}
    S_{\max}(t) = \frac{2}{N(t)} \sqrt{c^2 - s^2 + f^2} > 2,
\end{equation*}
where $c = e^{-\gamma_2 t}\cos(\omega_z t)$, $s= e^{-\gamma_2 t}\sin(\omega_z t)$ and $f = \text{tanh}^2\left( \frac{\hbar \omega_z}{2 k_B T} \right) + \text{sech}^2\left( \frac{\hbar \omega_z}{2 k_B T}\right)e^{-\gamma_1 t}$. Using the same parameters as in Fig. 1, Fig. 2 shows the time dependence of the maximised Bell parameter.

Initially, $S_{\max}$ satisfies the quantum Tsirelson bound - as expected, because this is when the qubit is in its most coherent state; gradually, as the system relaxes into a steady-state, $S_{\max}$ falls below the classical bound.

Interestingly, we note that, while $S_{\max}$ experiences the strongest violation at the earliest times, PDM negativity (see Fig. 1) only begins at $t = \pi$/4. The delayed PDM negativity in this Lindblad qubit model arises from the fact that temporal non-classicality (as diagnosed by eigenvalues of the PDM) depends on cross-time coherence building up in the measured basis, and this build-up doesn’t happen instantaneously in the open-system dynamics considered here. Negativity is essentially driven by the off-diagonal entries in the PDM, which encode coherent phase evolution between the two measurement times. Immediately after the first measurement, the system hasn’t had time to acquire a distinguishable quantum phase relative to $t_0$, and hence no negativity is observed. As time increases, the system picks up coherent oscillatory factors in the off-diagonal terms, that eventually dominate over the thermal diagonal terms to produce negativity in one of the eigenvalues. With further time evolution, decoherence (as described by the $e^{-\gamma_2 t}$ decay) suppresses the coherent terms and the negativity vanishes. 

Hence, while PDM negativity is a \textit{sufficient} condition for temporal non-classicality, it is \textit{not necessary}. A PDM with a negative eigenvalue cannot be reproduced by any classical, macrorealist model with non-invasive measurability, since classical temporal processes necessarily yield a positive semi-definite PDM \cite{fitzsimons2015quantum}.  This makes negativity a strong, basis-independent witness of temporal quantum correlations. However, it is not strictly necessary, because there exist temporal quantum processes where the PDM remains positive, but other indicators of temporal non-classicality - like temporal CHSH inequality violation - still show a quantum signature. The reason for this is that PDM negativity and temporal CHSH violation are sensitive to different features of the temporal correlations. PDM negativity is a statement about the entire two-time object $R$ being outside the set of classical states, independent of measurement choices and requires \textit{global} non-classicality. In contrast, the temporal CHSH inequality violation is a \textit{targeted} test that is basis-dependent, and probes the existence of a strong temporal correlation in some projection. It too is sufficient—but not necessary—for temporal entanglement, because maximal quantum violation only occurs in the optimal basis, and quantum processes can evade this criterion while satisfying other measures. 

Since these witnesses capture different facets of temporal correlations, a complete characterisation requires considering both in conjunction, as illustrated in Fig. 3. The figure shows the range of values of $C''(\omega)$, at each point in time, that produce temporal CHSH inequality violation for the PDM model. Even after $S_{\max}$ falls below the classical bound at $t \approx 2.5$ (see Fig. 2), values of the imaginary heat capacity persist in the quantum violation region until $t \approx 5.2$ - a timescale matching the survival of PDM negativity (see Fig. 1). Hence, even though both of these entanglement criteria are sufficient but not necessary when considered \textit{individually}, to fully characterise temporal non-classicality of the system it is necessary to consider both PDM negativity and temporal CHSH violation. 

This opens up an intriguing avenue for further exploration into the nature of temporal quantum correlations and the extent to which they may be probed via different entanglement measures. Moreover, it would be interesting to investigate the dependency of PDM negativity on the initial state of the qubit (\textit{i.e.} extend this work to consider pure and maximally-mixed initial states, for example). This would provide further insights into the ability of the PDM framework to faithfully characterise temporal non-classicality. Besides experimental validation of our predictions outlined in this paper (which is, of course, always the most reliable method), further corroboration could be achieved by considering two spatially separated qubits, each measured at two different times, and attempting to recover the results from the work by \textit{Wie$\acute{s}$niak et. al.} \cite{wiesniak2008heat} using the PDM framework. We leave these questions open for future investigations.

To conclude, we have presented a novel witness of temporal entanglement that is feasibly extendable to the macroscopic domain, and bypasses the experimental challenges currently posed when probing temporal Bell inequality violation at the emergent scale. We have provided a physical meaning, rooted in quantum information theory, to the imaginary component of the complex heat capacity, and have shown that it can be used to probe temporal non-classicality in the context of pseudo-density matrices and temporal CHSH inequalities. Further to this, we have placed a bound on the values of the imaginary heat capacity that serve as a direct witness of temporal entanglement, and will explore the experimental verification of these predictions in a forthcoming paper.

\textbf{Acknowledgements.} VV’s research is supported by the Gordon and Betty Moore and Templeton Foundations.
\bibliographystyle{apsrev4-1}
\bibliography{references}

\onecolumngrid
\section*{Appendix}
We consider an open system governed by the Lindblad master equation \cite{lindblad1976generators, breuer2002theory}, 
\begin{equation}
    \frac{d\rho}{dt} = \frac{-i}{\hbar} [H, \rho] + \sum_j \left( L_j \rho L_j^{\dagger} -\frac{1}{2} \{L_j^{\dagger} L_j, \rho \} \right)
\end{equation}
consisting of a qubit coupled to a thermal bath. The Hamiltonian of the qubit is given by $H = \frac{\hbar\omega_z}{2} \sigma_z$, and has eigenstates $|0\rangle$ and $|1\rangle$, with corresponding eigenvalues $\pm \frac{\hbar \omega_z}{2}$. The coupling of the qubit to the thermal bath may be modelled via the Lindblad jump operators $L_{-} = \sqrt{\gamma_{-}} \sigma_{-}$ and $L_{+} = \sqrt{\gamma_{+}} \sigma_{+}$, where (stimulated/spontaneous) emission and absorption are described by $\gamma_{-}$ and $\gamma_{+}$ respectively. If a temperature perturbation of the form $T(t) = T + \delta T e^{-i \omega t}$ is applied, the Hamiltonian of the qubit becomes $H = \frac{\hbar\omega_z}{2} \sigma_z + \delta H(t)$, where $\delta H(t) = - \left(\frac{\partial \langle H \rangle}{\partial t}\right) \delta T(t) = -C \delta T(t)$. We now proceed to solve the Lindblad equation, with a state of the general form $\rho(t) = \frac{1}{2} \left( \mathbf{1} + x(t) \sigma_x  + y(t) \sigma_y + z(t) \sigma_z\right)$.

For simplicity, here only $z(t) = Tr[\rho(t) \sigma_z]$ is of interest; later, we will generalise the Hamiltonian of the qubit to also include the off-diagonal terms and briefly discuss changes to the final result. From Eq.(10), we obtain $\dot{z}(t) = - \Gamma \left( z(t) - z_{eq}(t) \right)$, where $\Gamma = \gamma_- + \gamma_+$ and $z_{eq}(t)$ is the instantaneous thermal equilibrium magnetisation, given by $z_{eq} = -\tanh \left( \frac{\hbar \omega_z}{2 k_B T}\right)$. Linearising around small $\delta T(t)$, we can write $z_{eq}(t) \approx z_{eq}(0) + \delta T(t) \left(\frac{\partial z_{eq}}{\partial T}\right)$ and arrive at
\begin{equation}
    \dot{z}(t) + \Gamma z(t) = \Gamma \left( z_{eq}(0) + \delta T(t)\left( \frac{\partial z_{eq}}{\partial T}\right)\right).
\end{equation}
This ordinary differential equation is best solved in frequency space. Performing the following transformations $\delta T(t) = \delta T e^{-i \omega t}$ and $z(t) = z(0) + \delta z(\omega) e^{-i \omega t}$ leads to
\begin{equation}
    \delta z(\omega) = \frac{\Gamma}{\Gamma - i \omega} \delta T \left( \frac{\partial z_{eq}}{\partial T} \right).
\end{equation}
In frequency space, the complex heat capacity becomes $\delta \langle H(\omega) \rangle = C(\omega) \delta T(\omega)$ and noting that $\delta \langle H(\omega) \rangle = \frac{\hbar \omega_z}{2}\delta z(\omega)$, the imaginary component of the heat capacity is given by
\begin{equation}
    C''(\omega) = \left( \frac{(\hbar \omega_z)^2}{4 k_B T^2} \right)\frac{\omega \Gamma}{\Gamma^2 + \omega^2} \text{sech}^2\left( \frac{\hbar \omega_z}{2 k_B T}\right).
\end{equation} 
Generalising this result for a Hamiltonian that includes the off-diagonal terms $H = \frac{\hbar \omega}{2} (\hat{n} \cdot \vec{\sigma})$, we follow the same procedure as above. The eigenstates in this case are $|+\rangle$ and $|-\rangle$, with eigenvalues $\pm \frac{\hbar \omega}{2}$. The Lindblad jump operators take the same form as before, but note that the raising and lowering operators are now in the new eigenbasis. For ease, we can rotate the state $\rho \xrightarrow{} \tilde{\rho} = U^{\dagger} \rho U$, where the matrix $U$ performs a rotation from $\hat{z} \xrightarrow{} \hat{n}$, and work in the rotated frame for all subsequent calculations. The total heat capacity behaves as if the system were thermalised in the rotated eigenbasis of the Hamiltonian and the imaginary component becomes
\begin{equation}
    C''(\omega) = \left( \frac{(\hbar \omega)^2}{4k_B T^2}\right)\frac{\omega \Gamma}{\Gamma^2 +\omega^2}\text{sech}^2 \left(\frac{\hbar \omega}{2 k_B T}\right),
\end{equation}
which is the quantity used in the paper. 
\end{document}